\documentclass[twocolumn,prl,aps,showpacs,superscriptaddress,epsfig,dvips,amsmath,amssymb]{revtex4}
\usepackage{eufrak}
\usepackage[dvips]{graphicx}% Include figure files
\usepackage{dcolumn}% Align table columns on decimal point
\usepackage{bm}% bold math
\usepackage{multirow}
\newcommand{\nablav}{{\vec\nabla}}

\newcommand{\rv}{{\vec r}}
\newcommand{\Pv}{{\vec\Pi}}

\newcommand{\xv}{{\vec x}}
\newcommand{\Xv}{{\vec X}}

\newcommand{\uv}{{\vec u}}
\newcommand{\vv}{{\vec v}}

\newcommand{\qv}{{\vec q}}
\newcommand{\pv}{{\vec p}}

\newcommand{\Tr}{{\rm Tr}}

\newcommand{\zh}{{\hat{z}}}

\newcommand{\qh}{{\hat{q}}}

\newcommand{\mm}[1]{{\bf #1}}

\def\mm#1{ {\mathbf #1}}

\def\gm{{\mathbf{g}}}

\def\lm{{{\mathbf \lambda}}}
\def\Lm{{{\mathbf \Lambda}}}
\def\gm{{{\mathbf g}}}

\newcommand{\PL}{{{\mathbf P}^{\rm L}}}

\begin{document}
%\preprint{APS/123-QED}
\title{Thermal Fluctuations and Rubber Elasticity}
\date{\today}
\author{Xiangjun Xing}
\affiliation{Department of 
Physics, Syracuse University, Syracuse, New York~13244}
\email{xxing@physics.syr.edu}
\author{Paul M. Goldbart}
\affiliation{Department of Physics,
University of Illinois at Urbana-Champaign,
1110 West Green Street, Urbana, Illinois~61801-3080}
%Lines break automatically or can be forced with \\
\author{Leo Radzihovsky}
%\email{Second.Author@institution.edu}
\affiliation{Department of Physics, University of Colorado,
Boulder, Colorado 80309}%
\date{September 20, 2006}
% [CHECK]
\date{\today} %freeze this upon submission
\begin{abstract}
  The effects of thermal elastic fluctuations in rubber materials are
  examined.  It is shown that, due to an interplay with the
  incompressibility constraint, these fluctuations qualitatively
  modify the large-deformation stress-strain relation, compared to that of
  classical rubber elasticity.  To leading order, this mechanism
  provides a simple and generic explanation for the peak structure of
  Mooney-Rivlin stress-strain relation, and shows a good agreement with
  experiments.  It also leads to the prediction of a phonon correlation function that
  depends on the external deformation.
\end{abstract}

% [CHECK: need to add PACS numbers]
%\pacs{61.41.+e, 61.43.Er, 62.20.Dc, 64.60.Ak, 64.60.Fr, 64.70.Dv}% PACS, the
%61.41.+e 61.43.-j
%Physics and Astronomy
% Classification Scheme.
%\keywords{Suggested keywords}%Use showkeys class option if keyword
%display desired
\maketitle

\noindent
%\Begin{multicols}{2}
%\section{Introduction}
%label{Sec:Intro}

The term rubber (elastomer) refers to amorphous, essentially
incompressible, solids that consist of a crosslinked polymer network
and that can sustain large, reversible, shear deformations.  It has long
been understood that the elasticity of rubber is predominantly
entropic, being associated with the suppression of the entropy of the
polymer network by the imposed deformation. As a result rubber elasticity is
characterized by a shear modulus that is proportional to temperature.

The classical theory of rubber
elasticity~\cite{book:Treloar,ref:RCtext}, developed by Kuhn, Wall,
Flory, Treloar, and many others, around the 1940's, qualitatively
accounts for the entropic nature of rubber elasticity.  It is based on
the crucial assumption that the junctions of polymer networks do not
fluctuate in space, but nevertheless deform affinely with an imposed
uniform shear strain.  The entropy of the entire rubber network is
then given by the sum of the entropies of each polymer chain.  For a
uniform shear, i.e., for a homogeneous, volume-preserving deformation
$\Lm$, the elastic free-energy density $f$ is given by
\begin{equation}
f_0 = -\frac{k_B T}{V}\,\delta S = \frac{1}{2} \mu_0 \, \Tr \,\Lm^{\rm T} \Lm,
\label{classical-model}
\end{equation}
where $T$ is the temperature, $\delta S$ is the total entropy change
due to $\Lm$, $V$ is the volume, and $\mu_0\approx
k_B T/\xi^d$ is the entropic shear modulus in $d$ dimensions, with
$\xi$ being the typical mesh size of the polymer network.  For a {\it
  uniaxial\/} shear deformation along $\zh$,
\begin{equation}
\Lm = (\lm - \lm^{-1/2} )\, \zh\zh + \lm^{-1/2}\,\mm{I},
\label{uniaxial-deform}
\end{equation}
and the classical theory predicts
\begin{eqnarray}
f_0(\lambda) &=&
\frac{1}{2}\, \mu_0 \, \left( \lambda^2 + \frac{2}{\lambda} \right).
\label{fclassical}
\end{eqnarray}

\begin{figure}
\begin{center}
\includegraphics[width=9.5cm]{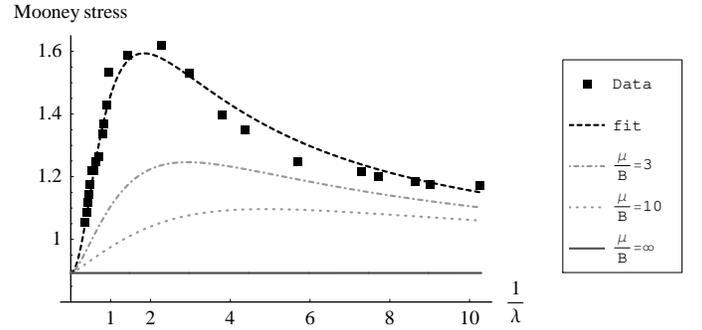}
\caption{The Mooney-Rivlin plot of the Mooney stress
  $({df}/{d\lambda})/(\lambda-\lambda^{-2})$ for a uniaxial shear
  deformation, Eq.~(\ref{uniaxial-deform}), versus $1/\lambda$.  While
  the classical theory predicts a horizontal line, real rubbery
  materials universally exhibit a pronounced peak feature.  Squares:
  data from Xu and Mark~\cite{Xu-Mark} (unit: $10^5$ Pa); Top curve: our theory fit that
  incorporates lowest-order phonon fluctuations on scales beyond
  crosslink spacing, using Eq.~(\ref{elastic-f-2}).  Fitting
  parameters: $B = \infty, \mu_0 = 8.92, \mu_1= 7.10$. 
   Also plotted are the curves for finite bulk moduli $B$,
  which are calculated using Eq.~(\ref{deltafB}) and same fitting
  parameters $\mu_0$ and $\mu_1$. }
\label{Mooney-Rivlin}
\end{center}
\vspace{-8mm}
\end{figure}

It has long been known that the classical theory does not work well
for large deformations~\cite{book:Treloar}. Its failure becomes most
salient in the so-called Mooney-Rivlin plot of the stress-strain relation,
in which $({df}/{d\lambda})/(\lambda-\lambda^{-2})$ is plotted versus
$1/\lambda$.  While from Eq.(\ref{fclassical}) it is clear that the
classical theory predicts a horizontal line at $\mu_0$ in such a plot,
almost all rubbery materials, natural or synthetic, exhibit universal
and nontrivial features (e.g., a peak around a compressed state), as
illustrated in Fig.\ref{Mooney-Rivlin}.

A number of mechanisms put forward to explain these features, including
polymer entanglement~\cite{Edwards-tube,Rubinstein-tube,Terentjev-tube},
non-Gaussian chain statistics, irreversible effects, internal energy
effects, as well as nematic order of various types, and
crystallization~\cite{book:Treloar,ref:RCtext}.  However, to date,
there is no broad consensus on the nature of the dominant mechanism
responsible for the deviation of rubber elasticity from the classical
theory.

Our main result, Eq.~(\ref{elastic-f-2}), plotted in
Fig.\ref{Mooney-Rivlin}, is a generic and simple explanation for the
Mooney-Rivlin data in terms of {\em incompressible} phonon
fluctuations of the rubber network.  It is based on the key
observation that the classical theory is an effective mean-field
theory that misses a large entropic contribution to the free energy,
due to thermal phonon fluctuations of the polymer network having
wavelengths longer than $\xi$.  These corrections are in fact {\em
  comparable} to the classical elastic free-energy density,
Eq.~(\ref{classical-model}), whose scale is given by the shear modulus
$\mu_0\approx k_BT/ \xi^d$. To see this, we note that, by the
equipartition theorem, each phonon mode contributes $k_B T$ to the
total free energy, and that the phonon density of the state is given by $\xi^{-d}$, set
by the mesh size $\xi$, which is the short-scale cutoff for these
long-wavelength fluctuations~\cite{commentSolids}.

It is noteworthy that these elastic fluctuations were previously
studied by James and Guth~\cite{James-Guth} in a {\it phantom\/}
network, where interactions between neighboring polymer chains are
ignored.  They concluded that these fluctuations are {\em independent} of
the imposed strain deformation and therefore have {\em no} effect on
the elasticity.

However, real rubber is nearly {\em incompressible}.  So, in addition
to the global volume preserving constraint on the macroscopic
distortion, i.e.~$\det\Lm=1$, the elastic fluctuations are constrained to
preserve the {\em local} density.  As we shall show, the interplay between
this local incompressibility constraint and an imposed macroscopic
shear deformation alters the spectrum of phonon fluctuations. This
leads to an entropic correction to the free energy that
depends strongly on the imposed distortion, and thereby qualitatively modifies
the stress-strain relation beyond that of the classical theory.

To go beyond classical rubber elasticity and incorporate thermal
fluctuations on scale beyond $\xi$, we consider the 
Lagrangian~\cite{commentPi} of an incompressible, homogeneous, 
isotropic elastic manifold (appropriate for an amorphous solid, when
local heterogeneities are ignored) with the elastic energy given by the
classical theory, Eq.~(\ref{classical-model}):
\begin{equation}
L = \int d^d\Xv \left[
\frac{1}{2} \rho\, \dot{\rv}(\Xv)\,^2
 - \frac{1}{2} \,\mu_0\, (\nablav\rv(\Xv))^2
\right], \label{Lagrangian}
\end{equation}
where $\rv(\Xv)$ is the elastic configuration of a $d$-dimensional
manifold satisfying the local incompressibility constraint
\begin{eqnarray}
\det \nablav \rv(\Xv)= 1, \label{constraint}
\end{eqnarray}
with $\Xv$ the equilibrium position of the mass point in the
undeformed reference state, and $\rho$ the mass density.  Now consider
a uniformly shear-strained state characterized by a deformation
gradient $\Lm$ that preserves the volume, i.e., $\det \Lm = 1$.  The
mass-point $\Xv$, which was fluctuating about the position $\Xv$ in
the unstrained state, now fluctuates around the new equilibrium position
$\Lm \cdot \Xv$ in the strained state.

Our task is therefore to sum over all long wave-length elastic
fluctuations in the strained state that satisfy local
incompressibility.  However, the highly {\em nonlinear} nature of the
constraint, Eq.~(\ref{constraint}), is the major impediment to the
analysis.  The key to our progress in the solution of this problem is
the resolution of the constraint (\ref{constraint}) through the following
parameterization of $\rv(\Xv)$:
\begin{equation}
\rv(\Xv) =  %\Lm \cdot \tilde{r}(\Xv) =
\Lm \cdot e^{\vec{v}(\Xv) \cdot \nablav} \Xv.
\label{r-v-2}
\end{equation}
It can be proved that $\rv(\Xv)$ is volume-preserving {\em if and only if} 
the field $\vec{v}(\Xv)$ is divergenceless, i.e.,
\begin{equation}
\nablav \cdot \vec{v}(\Xv) = 0,
\hspace{3mm}\mbox{or}\,\,\,\qv \cdot \vv_\qv= 0,
\label{constraint-2}
\end{equation}
where the second form is the equivalent transversality (to $\qv$)
condition on $\vv_\qv$, the Fourier transform of $\vv(\Xv)$.  It is
customary to call such a $\vv(\Xv)$ field an {\it incompressible
  flow\/}~\cite{Diff}. The resolution of the nonlinear constraint
Eq.~(\ref{constraint}) by the linear relation Eq.~(\ref{constraint-2})
in terms of $\vv$ paves the way for a systematic treatment of thermal
fluctuations in an incompressible elastomer.

Setting $\vv = 0$ in Eq.~(\ref{r-v-2}), we find $\rv_0 = \Lm\cdot
\Xv$, i.e. the uniformly strained reference state.  For small $\vv$,
the exponential in Eq.~(\ref{r-v-2}) is
\begin{equation}
\rv(\Xv) =  \Lm \cdot \left( \Xv + \vv(\Xv) 
 + \cdots \right),
\label{r-v-3}
\end{equation}
identifying $\Lm\cdot\vv(\Xv)$ with a phonon field displacement from
the uniformly-strained reference state $\rv_0$.

The canonical momentum field $\vec{\Pi}(\Xv)$ conjugate to $\vv(\Xv)$
is calculated in the standard way~\cite{commentPi}.  To lowest
order in $\vv$, we find
\begin{equation}
\vec{\Pi}(\Xv) = \frac{\delta L}{\delta \dot{\vec{v}}(\xv)} 
\approx \rho\, \gm \cdot \dot{\vv}(\Xv),
\label{Pi}
\end{equation}
where $\gm \equiv \Lm^{\rm T} \Lm $ is the metric tensor.

The Hamiltonian $H$ is related to the Lagrangian via the Legendre
transformation: $H[\vec{\Pi},\vec{v}] = \int d^d\Xv
\,\Pi(\Xv)\dot{\vec{v}}(\Xv) - L$. Using $\rv(\Xv)$, Eq.~(\ref{r-v-3}),
the Lagrangian Eq.~(\ref{Lagrangian}), and using Eq.~(\ref{Pi})
to eliminate $\dot{\vv}$ in favor of the conjugate momentum $\vec{\Pi}$, 
to lowest (i.e.~quadratic) order in $\vv$~\cite{commentPi}, we find
%\begin{widetext}
\begin{eqnarray}
H &=& E_0[\Lm] + \delta H_{\Lm} [\vec{\Pi},\vv], \\
\hspace{-1cm}
\delta H_{\Lm} 
&=& \int d\Xv \left[
\frac{1}{2\rho} \vec{\Pi}  \cdot \gm^{-1} \cdot \vec{\Pi}
+ \frac{\mu_0}{2} \partial_a \vv \cdot \gm \cdot
\partial_a \vv \right] .
\label{H_Lm}
\end{eqnarray}
Here, $E_0[\Lm]$ is the elastic energy for the uniform strained
reference state, identical to the classical theory result $f_0$,
Eq.~(\ref{classical-model}), whereas $\delta H_{\Lm}$ describes
collective elastic fluctuations, and depends explicitly on the uniform
shear deformation $\Lm$ through the metric tensor $\mm{g}$.

The partition function of a macroscopically sheared rubber is then
given by the following phase-space path integral:
\begin{eqnarray}
\hspace{-1cm}Z_{\Lm} &=&  
Z_{\Lm}^{\Pi} \cdot Z_{\Lm}^{v}=\int D\vec{\Pi} D\vv \prod_{\qv}
\delta(\qv \cdot \vv_\qv) \,e^{ - \beta\,\delta H_{\Lm}}, 
\label{Partition} 
\end{eqnarray}
where $\beta\equiv1/k_BT$.  We note that to quadratic order in $\vv$
and $\Pv$ (which is our focus here) the partition function,
Eq.~(\ref{Partition}), separates into a product of kinetic
($Z_{\Lm}^{\Pi}$) and elastic ($Z_{\Lm}^{v}$) parts.  Furthermore, the
incompressibility constraint only applies to the $\vv$ field but not to
the canonical momentum field $\Pv$.  As a result, the momentum
contribution $Z_\Lm^\Pi$ leads to an inconsequential
strain-independent constant. We emphasize, however, that because of
the {\em nonlinear} couplings between $\vv$ and $\Pv$, this property
does not persist to higher orders, and in our formulation (in terms of
$\vv$) the momentum degrees of freedom contribute nontrivially to
rubber elasticity~\cite{commentPi}.

The elastic part of the partition function is given by 
\begin{eqnarray}
Z_{\Lm}^v = \int D\vv \prod_{\qv}\delta(\qv\cdot \vv)
\,\exp\left[ -\frac{\beta\,\mu_0}{2}  \int_\qv \,q^2\,
\vv_\qv\cdot\gm\cdot\vv_{-\qv}\right].
\nonumber\\
\label{Z-v}
\end{eqnarray}
Because of the incompressibility constraint on $\vv$, encoded in the
$\delta$-functional, the dependence of the free energy on the imposed
strain $\Lm$ cannot simply be eliminated by a change of variables.
This contrasts with the aforementioned result of James and
Guth~\cite{James-Guth} for a phantom network, where such a constraint,
and concomitantly the dependence on the imposed deformation, are
absent.

The structure of $Z^v_\Lm$ for the unstrained case of $\mm{g} = \mm{I}$
is identical to that of a $U(1)$ gauge field theory in the transverse
gauge. Representing the $\delta$-functional by its functional Fourier
representation, $Z_{\Lm}^{v}$ is easily computed via two standard
Gaussian integrations:
\begin{eqnarray}
Z^v_{\Lm} &=& \int D\alpha_\qv D\vv_\qv
\,e^{-\int_\qv\left(\frac{\beta\,\mu_0}{2}q^2\,
\vv_\qv\cdot\gm\cdot\vv_{-\qv}+i\alpha_\qv\qv\cdot\vv_\qv\right)},
\nonumber\\
&=& \prod_\qv\left({2\pi\beta\mu_0\over\Tr(\PL_\qv\gm^{-1})}\right)^{1/2},
\label{Z-v-2}
\end{eqnarray}
where $\PL \equiv \qv\qv/q^2 = \qh\qh$ is the longitudinal projection
operator onto $\qh$.  Ignoring irrelevant $\Lm$-independent additive
constants, we obtain the free-energy correction due to elastic
fluctuations from scales longer than $\xi$:
\begin{eqnarray}
\delta F_\Lm &=& - k_BT\ln Z_{\Lm}^{v}
=\frac{1}{2}k_BT\sum_\qv\ln(\Tr\PL\gm^{-1}),\nonumber\\
&=& \frac{1}{2}k_BT V\xi^{-d}\Omega_d\,
\big\langle\ln(\Tr\PL\gm^{-1})\big\rangle_\qh,
 \label{dF-v}
\end{eqnarray}
where $\langle\cdots\rangle_{\qh}$ denotes an average over the
orientation of the $d$-dimensional unit-vector $\qh$, while $\Omega_d$
is a numerical factor of order of the surface area of a d-dimensional unit sphere.

By combining this result with the classical contribution,
$E_0[\Lm]=f_0[\Lm]$, Eq.~(\ref{classical-model}), 
we arrive at the central result of this
letter, i.e., the elastic free-energy density of an incompressible
rubber, subject to a uniform shear deformation $\Lm$, computed to
lowest-order in the thermal fluctuations:
\begin{equation}
f(\Lm) = \frac{\mu_0}{2}\, \Tr \,\gm +
\frac{1}{2}T\, \Omega_d \, \xi^{-d}\,
 \big\langle  \ln(\Tr\,\PL \gm^{-1})\big\rangle_{\qh}.
 \label{elastic-f}
\end{equation}
As was argued earlier, the scale of fluctuation contribution (i.e. the second term), 
measured by $T\, \Omega_d \, \xi^{-d}$, is of the same order of magnitude 
as the classical, mean-field contribution set by the shear modulus $\mu_0$.

For a three-dimensional uniaxially deformed system, with $\Lm$ given
by Eq.~(\ref{uniaxial-deform}), we have calculated the average in
Eq.~(\ref{elastic-f}) explicitly, and thus obtain an analytical expression
for $f(\lambda)$.  Ignoring a $\lambda$-independent constant, we find
\begin{eqnarray}
\hspace{-0.3cm}
f =  \frac{1}{2}\, \mu_0 \, (\lambda^2 + \frac{2}{\lambda})
+ \mu_1 \left[ 
 \frac{\tanh^{-1}\sqrt{1-\lm^{-3}}}{\sqrt{1-\lm^{-3}}}
 - \ln\lambda \right],
\label{elastic-f-2}
\end{eqnarray}
where $\mu_1 \equiv 2\, T\,\Omega_d \xi^{-d}$.  
For compression, i.e., $\lambda<1$, the
second term should be analytically continued in such a way that it
remains real and positive, namely with
$(1-\lambda^{-3})^{-1/2}\tanh^{-1}\sqrt{1-\lambda^{-3}}\rightarrow
(\lambda^{-3}-1)^{-1/2}\tan^{-1}\sqrt{\lambda^{-3}-1}$, for $\lambda <
1$.

In Fig.~\ref{Mooney-Rivlin} we compare our prediction for the
corresponding Mooney stress, $({df}/{d\lambda})/(\lambda-\lambda^{-2})$,
with that extracted from the stress-strain curve of Ref.~\cite{Xu-Mark},
and find excellent agreement.  The ``mean-field'' and long-wavelength
fluctuations shear moduli, $\mu_0$ and $\mu_1$, respectively provide
two independent fitting parameters, which only vertically translate and scale
the curve in the Mooney-Rivlin plot, Fig.~\ref{Mooney-Rivlin}.  However, 
the {\it shape\/} of the curve is completely determined by the
fluctuation contribution (i.e., second term) of the RHS of
Eq.~(\ref{elastic-f-2}), and thus has {\em no free parameters}.
Comparison of Eq.~(\ref{elastic-f}) with experimental data on biaxial
deformations (analysis of which we leave for the future) should provide
more stringent test of our theory.

Our analysis can be easily extended to a more realistic system with a
finite bulk modulus $B$.  This can be done by removing the hard
constraint, Eq.~(\ref{constraint}) and Eq.~(\ref{constraint-2}), and
adding a term $B (\nablav\cdot \vec{v})^2 / 2$ to the Hamiltonian,
Eq.~(\ref{H_Lm}), which suppresses density fluctuations.  In the case
of a uniaxial distortion of a compressible rubber, we find the lowest-order 
fluctuation correction to the free energy becomes
%\begin{widetext}
\begin{eqnarray}
\hspace{-0.3cm}
\delta f(\lambda,\tau) 
&=& \frac{\mu_1}{2}\bigg[\sqrt{1+\tau/\lambda\over 1-\lm^{-3}}
\tanh^{-1}\left(\sqrt{1-\lm^{-3}\over1+\tau/\lambda}\right)\nonumber\\
&&\,\,\,\,\,\,\,\,\,+\frac{1}{2}\ln(\lambda^{-2}+\tau)\bigg],
\label{deltafB}
\end{eqnarray}
%\end{widetext}
where $\tau \equiv \mu_0/B$.  As illustrated in Fig.~\ref{Mooney-Rivlin},
this result naturally interpolates between an incompressible and
phantom (classical) rubber.  This result is consistent with experiments that
observe a systematic reduction in the deviation from the classical theory
with softening of the bulk modulus via swelling, i.e. adding solvent
to the system~\footnote{See, for example, chapter 5 of reference
  \cite{book:Treloar} for a review. }.

The correlation function
\begin{equation}
V\, G^{ab}_\qv =
\langle
u^a_\qv\,
u^b_{-\qv}
\rangle=
\Lambda_{ai}\,\Lambda_{bj}\,
\langle v^i_\qv\, v^j_{-\qv} \rangle
\label{correlation}
\end{equation}
of the phonon field $\uv(\Xv) \approx \Lm \cdot \vv(\Xv)$ (defined by
Eq.~(\ref{r-v-3})), relative to a macroscopically strained state $\rv_0
= \Lm\cdot \Xv$, can also easily be computed. For an incompressible
rubber, a Gaussian integration with the Boltzmann weight from
Eq.~(\ref{Z-v-2}) gives
\begin{equation}
G^{ab}_\qv = {k_B T\over\mu_0 q^2}\left(\delta_{ab}-
{\Lm^{-1}_{ia} \Lm^{-1}_{jb}\qh_i\qh_j
\over\Tr\,\PL \gm^{-1}}\right).
\label{Gu-0}
\end{equation}

There is a small caveat, however.  We have been labeling the mass
points in the deformed state $\rv_0 = \Lm\cdot \Xv$ by their
undeformed equilibrium coordinate $\Xv$, which is conjugate to the
wave-vector $\qv$.  However, the wavevector $\pv$ probed by scattering
experiments is the one conjugate to the deformed equilibrium position
$\rv_0 = \Lm\cdot \Xv$.  These two vectors are related via
\begin{equation}
(\pv, \Lm \cdot \Xv) = (\qv, \Xv)
\longrightarrow \qv = \Lm^{\rm T} \cdot \pv,
\label{p-q}
\end{equation}
where $(\cdot,\cdot)$ denotes the inner product of two $d$-dimensional
vectors.  Therefore, the phonon correlation function, as a function of
the {\em physical} wave-vector $\pv$ is given by Eq.~(\ref{Gu-0}) with $\qv$
expressed in term of $\pv$ through Eq.~(\ref{p-q}).  This leads to
\begin{equation}
G^{ab}_\pv = \frac{T/\mu_0}{\pv\cdot\Lm\Lm^{\rm T}\cdot \pv} \,\,
\left( \delta_{ab} - \hat{p}_a \hat{p}_b \right) 
\propto P^{\rm T}_{ab}(\pv),
\end{equation}
which is proportional to the transverse projector, as expected from
incompressibility.  The predicted anisotropic strain-dependence
provides an independent test of the theory.

In this letter we have demonstrated the importance of thermal
fluctuations for the elasticity of isotropic rubber, particularly in the
large-deformation regime.  It is not difficult to see, however, that
the same general mechanism extends to all incompressible soft solids,
such as liquid crystalline elastomers.  Our current analysis is
limited to the lowest order.  Because the effective coupling constant
is of order unity, we expect high order corrections to be
quantitatively important.  However, as illustrated in
Fig.\ref{Mooney-Rivlin}, the lowest-order contributions already
capture the essential effects of thermal fluctuations on rubber
elasticity.  Our formalism provides a systematic approach for addressing
these high order contributions.  We hope that our work will
stimulate further studies in this direction.

\begin{acknowledgments}
  We acknowledge the hospitality of Boulder Summer School and the
  Aspen Center for Physics, where part of this work was done.  We
  thank N. Goldenfeld, P. M. Chaikin, and especially T. C. Lubensky for
  critical comments and discussions, and M.  Rubinstein for data
  reproduced in Fig.~\ref{Mooney-Rivlin}.  XX is especially indebted
  to M.  Rubinstein for his lectures at the 2002 Boulder Summer
  School, and to A.~P.~Balachandran and R. Sorkin for illuminating
  discussions on the mathematical formalism used in this work.  The
  authors acknowledge support from ACS under grant PRF 44689-G7 (XX),
  NSF under grants DMR 02-05858 and 06-05816 (PMG), and MRSEC
  DMR-0213918 and DMR-0321848 (LR).
\end{acknowledgments}

%\appendix`
%\section{Appendixes}

%\bibliographystyle{unsrt}     % or "siam", or "alpha", or "abbrv"
                               % see other styles (.bst files) in
                                %0 $TEXHOME/texmf/bibtex/bst
%\nocite{*}            % list all refs in database, cited or not.
%-------------
%\bibliography{reference}            % bib database file refs.bib

\vspace{-5mm}

\begin{thebibliography}{10}

\bibitem{book:Treloar}
L. R. G. Treloar,
{\sl The Physics of Rubber Elasticity\/}
(Clarendon Press, Oxford, 1975).

\bibitem{ref:RCtext}
M. Rubinstein and R. H. Colby,
{\sl Polymer Physics\/}
(Oxford University Press, 2003).

%\bibitem{deGennes-polymer}
%P. G. de Gennes,
%{\it Scaling Concepts in Polymer Physics\/}
%(Cornell University Press, Ithaca, NY, 1979).


%\bibitem{Deam-Edwards}
%R. T. Deam and S. F. Edwards,
%Phil. Trans. R. Soc. {\bf A, 280}, 317 (1976).

%\bibitem{Alexander-PR}
%S.~Alexander, Physics Report {\bf 296} (1998) 65-236
%% Amorhhous solids, their structure, lattice dynamics and elasticity
%% An extensive discussion of random solids, has lots of insights,
%% but few definite results.

%\bibitem{ref:Peng+Goldbart2000}
%W. Peng and P. M. Goldbart,
%Phys. Rev. E {\bf 61\/}, 3339 (2000).
% Renormalization-group approach to the vulcanization transition
% Phys. Rev. E {\bf 61\/}, 3339-3357 (2000).

%\bibitem{ref:Peng+Goldbart+McKane2001}
%H.-K. Janssen and O. Stenull,
%Phys. Rev. E {\bf 64\/}, 026119 (2001);
% Connecting the vulcanization transition to percolation
% Physical Review E 64, 031105 (2001), 7 pages
%W. Peng, P. M. Goldbart and A. J. McKane,
%Phys. Rev. E {\bf 64\/}, 031105 (2001).
% ...because they describe fluctuation effects

%\bibitem{Pak-Flory}
%H. Pak, P. J. Flory,
%J. Polym. Phys. {\bf 17}, 1845 (1979).

\bibitem{Xu-Mark} P. Xu and J.E. Mark, Rubber Chem. Technol, {\bf 63},
  276 (1990).  We thank M. Rubinstein for providing this data.

\bibitem{Edwards-tube}
S. F. Edwards and T. A. Vilgis,
Rep. Prog. Phys. {\bf 51\/}, 243-297 (1988).
% The tube model theory of rubber elasticity

\bibitem{Rubinstein-tube}
M. Rubinstein and S. Panyukov,
Macromolecules, {\bf 35\/}, 6670 (2002).

\bibitem{Terentjev-tube}
S. Kutter and E. M. Terentjev,
cond-mat/0106371; and
Eur. Phy. J E {\bf 8\/}, 539 (2002).

\bibitem{commentSolids} In an ordinary solid, the free-energy
  contribution due to long-wavelength fluctuations is also $O(T/\xi^d)$,
  but in contrast (at low $T$, below the solid's melting temperature) is
  much smaller than the elastic part set by the elastic constants that
  have {\em energetic} origin.~\cite{Fisher-Wigner}.

%\bibitem{comment_entropy}
%This necessarily excludes systems characterized by rigidity
%percolation, as well as networks with bond-bending forces.

%\bibitem{RMP-fracton}
%See, e.g.,
%T. Nakayama {\it et al}., \rmp {\bf 66}, 381 (1994),
%and references therein.

%\bibitem{Thorpe-Tang}
%W. Tang and M. F. Thorpe,
%\prb {\bf 36}, 3798 (1987).

\bibitem{James-Guth}
H. M. James and E. Guth,
J. Chem. Phys. {\bf 11}, 455 (1943).

\bibitem{commentPi} Although it is perhaps surprising for this
  classical (i.e.~non-quantal) statistical mechanics problem (where one
  would expect kinetic degrees of freedom to inconsequentially
  decouple), a full phase-space formulation is convenient here. In it,
  the functional integration measure $D\Pv_\rv D\rv$ is invariant under
  canonical transformations, thereby allowing us to perform the nonlinear
  transformation to $\vv$, Eq.(\ref{r-v-2}) (required to resolve the
  incompressibility constraint), without introducing a highly nonlinear
  (in $\vv$) functional Jacobian. The price for this is a conjugate
  momentum, $\Pv_\vv$, that nonlinearly couples to $\vv$, and
  therefore cannot in general be integrated out.

\bibitem{Diff} The function $\rv(\Xv)$ is a volume-preserving
  automorphism of three-dimensional Euclidean space $E(3)$.  The
  collection of all such functions forms the diffeomorphism group
  Diff[E(3)], an infinite-dimensional non-Abelian, non-compact, Lie
  group.  The mathematical implication of Eq.~(\ref{r-v-2}) is that
  the vector field $\vec{v}(\Xv)$ is an element of the corresponding
  infinite-dimensional Lie algebra, $diff[E(3)]$. For an arbitrary,
  volume-preserving $\rv(\Xv)$, there may not exist a vector field
  $\vec{v}(\Xv)$ such that Eq.~(\ref{r-v-2}) holds.  However, one can
  always find such a $\vec{v}(\Xv)$ if $\vec{r}(\Xv)$ is sufficiently
  close to $\Xv$.  At relatively low temperatures, therefore, we
  expect that the summation over all volume-preserving deformations
  $\rv(\Xv)$ can be replaced by one over all incompressible flows
  $\vv(\Xv)$.
  
\bibitem{Fisher-Wigner}
D. S. Fisher,
Phys. Rev. B 26, 5009 (1982).
% Phys. Rev. B 26, 5009–5021 (1982)
% Shear moduli and melting temperatures of two-dimensional electron crystals:
% Low temperatures and high magnetic fields

%\bibitem{comment_replicalimit}
%This limit is necessary in order to model the quenched
%nature of the crosslinks; see Ref.\cite{Deam-Edwards}
%for a detailed discussion.  Before taking the replica
%limit we treat $n$ as a positive integer.

%\bibitem{vulcan_Goldbart}
%P.~M.~Goldbart {\it et al.\/}, Adv. Phys. {\bf 45}, 393 (1996).

%\bibitem{XMG:scaling}
%X.~Xing, S.~Mukhopadhyay, P.~M.~Goldbart,
%Phys. Rev. Lett. {\bf 93}, 225701 (2003).

%\bibitem{XMG:unpublished}
%X.~Xing, S.~Mukhopadhyay, P.~M.~Goldbart, to be published.

%\bibitem{XMGZ:nematic}
%X.~Xing, S.~Mukhopadhyay, P.~M.~Goldbart, A. Zippelius,
%cond-mat/0411660.

%\bibitem{nonaffine}
%T.~ H$\ddot{o}$lzl, {\it et al},
%Phys. Rev. Lett. {\bf 79}, 2293 (1997);
%L.P.~Wittmer, {\it et al}, Europhys. Lett., {\bf 57}, 423 (2002);
%D.A.~Head, {\it et al}, {\bf 91}, 108102 (2003).

%\bibitem{Leibler:butterfly}
%J.~Bastide and L.~Leibler, Macromolecules,
%{\bf 21}, 2649 (1988);
%J.~Bastide, L.~Leibler, and J.~Prost, Macromolecules,
%{\bf 23}, 1821 (1990)

%\bibitem{Onuki:butterfly}
%A.~Onuki, J. Phys. II France, {\bf 2}, 45 (1992).

%\bibitem{Exp:butterfly}
%E.~Menders {\it et al}, Phys. Rev. Lett. {66}, 1595 (1991);
%E.~Straube {\it et al}, Phys. Rev. Lett. {74}, 4464 (1995).

%-------------
\end{thebibliography}
%-------------

%-------------
%\end{multicols}
\end{document}